\documentclass [12pt]{article}
\pdfoutput=1

\usepackage{amsmath}
\usepackage{amsfonts}
\usepackage{amscd}
\usepackage{amsthm}
\usepackage{setspace}

\usepackage{graphicx}
\usepackage{authblk}
\usepackage{caption}
\usepackage{ytableau}
\usepackage{mathtools}
\def\til{\tilde}

\begin{document}
	
\begin{titlepage}
		
\begin{flushright}
KEK-TH-2329			
\end{flushright}
		
\vskip 3.0cm
		
\begin{center}
			
{\large Path integrals in JT gravity and Virasoro constraints}
			
\vskip 1.2cm
			
Yusuke Kimura$^1$ 
\vskip 0.6cm
{\it $^1$KEK Theory Center, Institute of Particle and Nuclear Studies, KEK, \\ 1-1 Oho, Tsukuba, Ibaraki 305-0801, Japan}
\vskip 0.4cm
E-mail: kimurayu@post.kek.jp
			
\vskip 2cm
\abstract{We examine the large-$g$ asymptotic Weil--Petersson volume formulas deduced in the previous literature. The volume formulas have application to computing the partition functions and the correlation functions in Jackiw--Teitelboim gravity. We utilize two approaches to assess the validity of the formulas. The first approach is to examine the asymptotic volume formulas from the perspective of the Witten conjecture. If the volume formulas are correct, the generating function of the intersection indices deduced from the asymptotic volume formulas would satisfy constraints that are analogous to the Virasoro constraints. We confirmed that the intersection indices computed from the large-$g$ asymptotic volume formulas satisfy variants of the string and dilaton equations. This implies that the generating function of the intersection indices deduced from the asymptotic volume formulas indeed satisfies constraints analogous to first two of the Virasoro constraints. As another approach, we also examined the asymptotic volume formulas by studying the behavior of the higher-order spectral form factors. Our analyses suggest that the large-$g$ asymptotic Weil--Petersson volume formulas yield plausible estimations.}  
			
\end{center}
\end{titlepage}
	
\tableofcontents
\section{Introduction}
\par Jackiw--Teitelboim (JT) gravity \cite{Teitelboim83, Jackiw84} is a two-dimensional (2D) topological quantum gravitational theory with a dilaton that is considered on Riemann surfaces, possibly with boundaries. When the Riemann surfaces have asymptotic boundaries, the quantum fluctuations or ``wiggles'' along the asymptotic boundaries are controlled by the Schwarzian theory \cite{Kitaevtalk, MS2016, Kitaev2017}. As the one-dimensional Schwarzian theory describes the low-energy limit of the Sachdev--Ye--Kitaev (SYK) model \cite{Sachdev92, Kitaevtalk, Kitaev2017}, it relates JT gravity to SYK models. Moreover, JT gravity has been studied in the context of AdS/CFT correspondence \cite{Almheiri2014}. 
\par Recently, the duality of JT gravity and the matrix integral has been discussed in \cite{SSS2019}. There is a matched correspondence between two different recursion relations that hold in JT gravity and matrix integral via the proposed duality, and this is an important feature of the proposed duality. Mirzakhani's recursion relation \cite{Mirzakhani2007} that computes the volume of the moduli of hyperbolic Riemann surfaces on the JT gravity side corresponds \cite{Eynard2007} to the Eynard--Orantin topological recursion \cite{Eynard2007tp} \footnote{The genus expansion of a Hermitian-matrix integral \cite{Eynard2004} is obtained via the topological recursion of Eynard and Orantin.} on the matrix-integral side; this correspondence is important in the discussion \cite{SSS2019} of the duality of the two theories. Recent studies of JT gravity include \cite{Blommaert2019, SSS2019, Cotler2019, Moitra2019, StanfordWitten2019, Okuyama2019, Almheiri2019, Marolf2020, OkuyamaSakai2020multi, Maxfield202006, Kimura2008, Momeni202009, Momeni2020, Alishahiha2020, Narayan2020, Moitra2021, Griguolo2021, Griguolo202106}. 

\vspace{5mm}

\par The path integral in the JT gravity on Riemann surfaces with asymptotic boundaries is obtained by computing the path integral over the wiggles along the asymptotic boundaries of the surfaces \cite{Jensen2016, MSY2016, EMV2016}, together with the path integral over the moduli of the Riemann surfaces with the geodesic boundaries and the path integral over the ``trumpets'' connecting an asymptotic boundary (with which a boundary wiggle is associated) and a geodesic boundary \cite{SSS2019}. The volume of the moduli of hyperbolic Riemann surfaces with geodesic boundaries is known as the ``Weil--Petersson volume.'' 
\par Topology of Riemann surface is determined by the genus ($g$), and the number ($n$) of the boundaries. We use the notation $\mathcal{M}_{g,n}$ to denote the moduli space of the Riemann surfaces of genus $g$ with $n$ boundaries. Computing the Weil--Petersson volume of the moduli space $\mathcal{M}_{g,n}$ of Riemann surfaces was known to be a difficult problem in mathematics. Mirzakhani discovered a method to compute the Weil--Petersson volumes, $V_{g,n}$, for any genus, $g$, recursively \cite{Mirzakhani2007}.
\par The path integral over a ``trumpet'' connecting an asymptotic boundary and a geodesic boundary of a Riemann surface was computed in \cite{SSS2019}. Thus, the correlation function in JT gravity is obtained when the Weil--Petersson volumes are successfully computed. Although the computation of the Weil--Petersson volumes for {\it any} genus ($g$) and for any number ($n$) of the geodesic boundaries is possible in principle, it is not simple.
\par Although the evaluation of the Weil--Petersson volume is, in principle, possible for any genus $g$ by utilizing Mirzakhani's recursion formula \cite{Mirzakhani2007} iteratively, an expression of the Weil--Petersson volume, $V_{g,n}(b_1, \ldots, b_n)$, cannot be obtained as a function of genus $g$ and the $n$ number of the geodesic boundaries. In the notation of of the Weil--Petersson volume, $b_1, \ldots, b_n$ denote the lengths of the geodesic boundaries. In fact, as genus $g$ becomes large, performing the actual computation becomes difficult. Obtaining the expression of the Weil--Petersson volume as a function of $g$ and $n$ is generally not considered a straightforward problem \footnote{Several efforts have been made to obtain the asymptotic expressions \cite{Penner92, Zograf2008, Kimura2008} of the Weil--Petersson volumes, $V_{g,n}$, as functions of $g$ when $g$ is large.}. 
\par The asymptotic expressions for the Weil--Petersson volumes, $V_{g,n}(b_1, \ldots, b_n)$, as functions of $g$ and $n$ were obtained when $g$ was large in \cite{Kimura2008} by utilizing partial differential equations \cite{DoNorbury, Do2008} that hold for the moduli of the hyperbolic Riemann surfaces. The use of the expressions of the Weil--Petersson volumes in \cite{Kimura2008} enables the direct evaluation of the higher-genus contributions to the correlation functions in JT gravity, avoiding the iterative steps to recursively compute genus-$g$ partition functions, which becomes difficult once $g$ becomes large.   

\vspace{5mm}

\par The aim of this study is to examine the validity of the large-$g$ asymptotic formulas of the Weil--Petersson volumes estimated in \cite{Kimura2008} from two different perspectives. As we mentioned previously, the evaluations of the Weil--Petersson volumes can be used to compute the correlation functions and partition functions in JT gravity. The large-$g$ asymptotic Weil--Petersson volumes deduced in \cite{Kimura2008} enable us to yield the higher-genus contributions to the correlation functions and partition functions. Proving the formulas obtained in \cite{Kimura2008} rigorously seems considerably difficult, however, the validity of the formulas can be examined, to some degree, in at least two different ways. 
\par One method is to analyze the large-g asymptotic formulas of the Weil--Petersson volumes utilizing methods used in the mathematical formulation of the Witten conjecture. Another method is to investigate the behaviors of the higher-order spectral form factors (SFFs). Higher-genus contributions to the correlation functions and partition functions can be computed as Laplace transforms of the large-$g$ asymptotics of the Weil--Petersson volumes. One can deduce the higher-order SFFs from these results. Higher-order SFFs in JT gravity was studied in a different approach in \cite{Mertens2020}. We study whether the behaviors of the higher-order SFFs that we deduce from the large-$g$ asymptotics of the Weil--Petersson volumes in \cite{Kimura2008} agree with those obtained in \cite{Mertens2020}. In brief, there are two themes in this study:
\begin{itemize}
\item[] 1. We examine whether the large-$g$ asymptotics of the Weil--Petersson volumes in \cite{Kimura2008} are valid from the perspective of the Witten conjecture. \\

\item[] 2. We examine the validity of the large-$g$ asymptotics of the Weil--Petersson volumes by studying the behavior of the higher-order SFFs that we compute from the higher-genus contributions to the correlation functions that are Laplace transforms of the the large-$g$ asymptotic of the Weil--Petersson volumes.

\end{itemize}

\vspace{5mm}

\par The equivalence of the two versions of quantum gravitational theories, JT gravity and matrix integral, in fact can be viewed as a variant of the Witten conjecture \cite{Witten1990}. The Witten conjecture \cite{Witten1990} was used for the equivalence of two models of 2D quantum gravitational theory, and the statement of the conjecture involved the intersection theory of the line bundles on the moduli of the Riemann surfaces and a matrix theory \footnote{The Virasoro conjecture proposed by the authors in \cite{Eguchi1997} yields a generalization of the Witten conjecture.}. A proof of the Witten conjecture was given in \cite{Kontsevich1992}. Proofs of this conjecture can also be found in \cite{Okounkov2000, Mirzakhani2007int, Kazarian2007}. Mirzakhani's method to prove the Witten conjecture in \cite{Mirzakhani2007int} used Mirzakhani's recursion formula \cite{Mirzakhani2007}. By proving the Witten conjecture, Mirzakhani \cite{Mirzakhani2007int} expressed the coefficients of the Weil--Petersson volume as intersections numbers of line bundles on the moduli space of the hyperbolic Riemann surfaces. Mathematical formulation of the Witten conjecture is that the partition function of the formal function defined from the intersection indices of line bundles on the moduli space of the Riemann surfaces satisfies the Virasoro constraints. We will explain this formulation in detail in section \ref{subsec2.1}. 

\par As we stated earlier, precisely expressing the Weil--Petersson volume, $V_{g,n}$, as a function of genus $g$ and the number asymptotic boundaries ($n$) is presently considerably difficult. However, when the discussion is limited to the region where genus $g$ of the Riemann surface is large, the Weil--Petersson volume, $V_{g,n}$, can be expressed as a function of $g$ and $n$, as obtained in \cite{Kimura2008}.  
\par Under conditions $g>>b_i$, $i=1, \ldots, n$, and $g>>1$, where $b_i$s denote the lengths of the geodesic boundaries of a Riemann surface, the Weil--Petersson volume, $V_{g,n}(b_1, \ldots, b_n)$, of the moduli of the Riemann surface with genus $g$ and $n$ geodesic boundaries was evaluated to have the following expression to the leading order \cite{Kimura2008} \footnote{For $n=1$, the expression of the Weil--Petersson volume, $V_{g,1}(b)$, was obtained in \cite{SSS2019} under conditions $g >> b$ and $g >> 1$.} : 
\begin{equation}
\label{asymptoticVgn in 2.1}
V_{g,n}(b_1, \ldots, b_n) \sim \sqrt{\frac{2}{\pi}} 2^n \, (4\pi^2)^{2g+n-3} \, \Gamma(2g+n-\frac{5}{2})\, \cdot \prod_{i=1}^n \frac{{\rm sinh}(\frac{b_i}{2})}{b_i}.
\end{equation}
\par However, the large-$g$ asymptotics (\ref{asymptoticVgn in 2.1}) do not appear simple, and it is not evident whether the asymptotic volumes (\ref{asymptoticVgn in 2.1}) yield the correct approximations. In this note, we assess the validity of the large-$g$ asymptotic formulas by applying some techniques in the formulation of the Witten conjecture. This approach is one of the novelties of this study, and this approach is not discussed in \cite{Kimura2008}. We present an evidence that the asymptotic formulas are plausible from a perspective different from that in \cite{Kimura2008}. 
\par One can compute the asymptotic intersection indices of line bundles on the moduli space of the Riemann surfaces from the asymptotic volume (\ref{asymptoticVgn in 2.1}) as we will discuss in section \ref{subsec2.2}. We will refer to the generating function of the asymptotic intersection indices as the ``asymptotic generating function'' in this study. Because the calculation of the asymptotic intersection indices is an approximation, it is unlikely that the asymptotic generating function satisfies the exact Virasoro constraints. However, if the formulas (\ref{asymptoticVgn in 2.1}) in \cite{Kimura2008} yields a valid approximation, it is reasonable to expect that the asymptotic generating function still satisfies a variant of the Virasoro constraints. We will provide an evidence in section \ref{subsec2.2} that the asymptotic generating function possesses this type of symmetry. 
\par The Virasoro constraints involve countably-infinitely many partial differential operators $L_n$, and the first two constraints are equivalent to the string and dilaton equations. Concretely, we confirmed that the asymptotic generating function satisfies variants of the two equations. This result suggests that the asymptotic generating function satisfies a variant of the Virasoro constraints.

\vspace{5mm}

\par As another theme, we will investigate the behaviors of the higher-order SFFs from the large $g$ asymptotic volumes (\ref{asymptoticVgn in 2.1}). We will find in section \ref{sec3} that our result agrees with that deduced in \cite{Mertens2020} with a different approach. This provides another supporting evidence for the asymptotic volume formulas (\ref{asymptoticVgn in 2.1}). Studies of the spectral form factor in JT gravity can be found, e.g., in \cite{Cotler2016, Saad201806, SSS2019, Saad201910, Mertens2020}.

\vspace{5mm}
   
\par We also estimate the higher-genus contributions to the resolvent on the matrix-integral side utilizing large-$g$ asymptotics of the Weil--Petersson volumes in \cite{Kimura2008} in appendix \ref{appendixa}. The computation results of the resolvents in this study might be a useful tool for confirming the duality of JT gravity and the matrix integral as discussed in \cite{SSS2019}. 

\vspace{5mm}

\par The intersection numbers of the line bundles on the moduli of the Riemann surfaces were discussed in the context of topological gravity in \cite{Witten1990, Dijkgraaf2018}.
\par The intersection numbers with one boundary were explicitly computed when the genus is large in \cite{Okuyama2019}, and correlation functions with two boundaries were explicitly computed in the low-temperature limit in \cite{OkuyamaSakai2020multi}. The authors in \cite{Okuyama2019, OkuyamaSakai2020multi} used the Korteweg--de Vries (KdV) hierarchy approach to calculate these; this is considerably different from the approach used in this study.

\vspace{5mm}

\par The remainder of this note is structured as follows. The main and novel results are states in sections \ref{subsec2.2} and \ref{sec3}.
\par We review Witten conjecture and the Virasoro constraints in section \ref{subsec2.1}. We will also mention the relation of the string and dilaton equations to the Virasoro constraints. 
\par In section \ref{subsec2.2}, we deduce the large-$g$ asymptotic intersection numbers of line bundles on the moduli space of the Riemann surfaces from the large-$g$ asymptotic Weil--Petersson volumes. We then demonstrate that the asymptotic generating function of the intersection indices obtained from the deduced large-$g$ asymptotic intersection numbers satisfies variants of the string and dilaton equations. This corresponds to the first theme of this study. 
\par We will study the behaviors of the higher-order SFFs by computing the correlation functions as the Laplace transforms of the asymptotic volumes (\ref{asymptoticVgn in 2.1}) in section \ref{sec3}, and this corresponds to the second theme.
\par In section \ref{sec4}, we present our concluding remarks. In appendix \ref{appendixa}, we also evaluate the resolvent on the matrix-integral side, which may provide a tool for confirming the duality \cite{SSS2019} of JT gravity and the matrix integral.

\section{Large-$g$ asymptotic Weil--Petersson volume and Witten conjecture}
\label{sec2}
\subsection{Review of Witten conjecture and Virasoro constraints as preliminaries}
\label{subsec2.1}
Here, we briefly review the Witten conjecture and Virasoro constraints. We will also mention the relation of the string and dilaton equations to the Virasoro constraints. The string and dilaton equations are equivalent to two of the Virasoro constraints. We will need these notions for the discussion in section \ref{subsec2.2} to examine the validity of the large-$g$ asymptotic Weil--Petersson volumes (\ref{asymptoticVgn in 2.1}).
\par The Witten conjecture \cite{Witten1990} is used to determine the equivalence of two versions of 2D quantum gravitational theory.  
\par Several proofs of the Witten conjecture have been provided. Kontsevich provided a proof of the Witten conjecture in \cite{Kontsevich1992}. Mirzakhani proved the Witten conjecture using a different approach in \cite{Mirzakhani2007int}. The recursion relation of the Weil--Petersson volumes was effectively used in the proof in \cite{Mirzakhani2007int}, and as proved in \cite{Mirzakhani2007}, which revealed that the Weil--Petersson volume of the moduli space of the Riemann surface with boundaries can be expressed as a polynomial in the geodesic boundary lengths $b_i$. The main aspect in the proof of the Witten conjecture given in \cite{Mirzakhani2007int} was to express the coefficients of the Weil--Petersson volume as a polynomial in $b_i$'s using the intersection numbers of the line bundles associated with the cotangent spaces to the marked points of the Riemann surface. Using these expressions, Mirzakhani revealed \cite{Mirzakhani2007int} that the partition function $exp(F)$ of the generating function $F$ of the intersection indices satisfy the Virasoro constraints as follows:
\begin{equation}
\label{eq Virasoro constraints}
L_n \, exp(F)=0. 
\end{equation}
\par The generating function $F$ is defined as follows: one considers the formal sum of the intersection indices, $F_g$:
\begin{equation}
\label{formal sum Fg in 3}
F_g(t_0, t_1, \ldots) = \sum_{\{d_i\}} \, \prod_{l>0} \frac{t_l^{n_l}}{n_l!} <\prod \tau_{d_i}>_g.
\end{equation}
The summation is performed over all sequences of non-negative integers, $\{ d_i \}$, with finitely many nonzero terms. $n_l$ in the formal sum (\ref{formal sum Fg in 3}) denotes the number of indices $i$ of $d_i$ such that $d_i=l$. Subsequently, the generating function $F$ is defined as:
\begin{equation}
\label{equation generating function}
F=\sum_{g=0}^{\infty} F_g \zeta^{2g-2}.
\end{equation}
\par As we mentioned previously, by expressing the coefficients of the Weil--Petersson volumes in $b_i$ in terms of the intersection numbers of the line bundles over the moduli space of the hyperbolic Riemann surfaces, Mirzakhani \cite{Mirzakhani2007int} revealed that the partition function $exp(F)$ satisfies the Virasoro constraints: $L_n exp(F)=0$, $n\ge -1$. 
\par The sequence of the differential operators, $L_{-1}$, $L_0$, $L_1$, $\ldots, L_n, \ldots$, is defined as follows:
\begin{eqnarray}
\label{Virasoroop in 3}
L_{-1} = & \partial_{t_0}+\sum_{i=1}^{\infty} t_{i+1} \partial_{t_i} + \frac{\zeta^{-2}}{2}t_0^2 \\ \nonumber
L_0 = & \sum_{i=1}^{\infty} \frac{2i+1}{2}\partial_{t_i}+ \frac{3}{2}\partial_{t_1}+\frac{1}{16} \\ \nonumber
L_n = & \sum_{i=0}^{\infty} \frac{(2n+2i+1)!!}{2^{n+1}(2i-1)!!}t_i\partial_{t_{n+i}}-\frac{(2n+3)!!}{2^{n+1}}\partial_{t_{n+1}} +\frac{\zeta^2}{2}\sum_{i=0}^{n-1} \frac{(2i+1)!!(2n-2i-1)!!}{2^{n+1}}\partial_{t_i}\partial_{t_{n-1-i}} \hspace{2mm} (n\ge 1). 
\end{eqnarray}
In (\ref{Virasoroop in 3}), $(2i+1)!!=(2i+1)\cdot (2i-1)\cdot \ldots 3\cdot 1$. The operators $L_n$, $n\ge -1$, are referred to as Virasoro operators. They satisfy the following relationship:
\begin{equation}
[L_n, L_m]=(n-m) L_{n+m}.
\end{equation}
\par We now mention the relation of two equations, the string and dilaton equations, to the Virasoro constraints (\ref{eq Virasoro constraints}). First, we introduce two equations, namely string and dilaton equations. We use $\mathcal{L}_i$ to represent the cotangent spaces to the marked points in a closed Reimann surface of genus $g$. We use $\psi_i$ to denote the first Chern class of the cotangent space, $\mathcal{L}_i$, namely $\psi_i=c_1(\mathcal{L}_i)$. We use $<\tau_{d_1}, \ldots, \tau_{d_n}>_g$ to represent the intersection number of powers of the line bundles $\psi_i$ as follows:
\begin{equation}
<\tau_{d_1}, \ldots, \tau_{d_n}>_g = \int_{\overline{\mathcal{M}}_{g,n}} \psi_1^{d_1} \ldots \psi_n^{d_n}.
\end{equation}
$d_i$ are non-negative integers and satisfy the relation $\sum_{i=1}^n d_i=3g-3+n$.
\par The intersection indices $<\tau_1^{d_i} \cdots \tau_n^{d_n}>$ satisfy the string and dilaton equations \cite{Witten1990} as follows:
\begin{eqnarray}
\label{strdileq in 3}
<\tau_0, \tau_{d_1}, \ldots, \tau_{d_n}>_g = & \sum_{d_i\ne 0} <\tau_{d_1}, \ldots, \tau_{d_i-1}, \ldots>_g, \\ \nonumber
<\tau_1, \tau_{d_1}, \ldots, \tau_{d_n}>_g = & (2g-2+n) <\tau_{d_1}, \ldots, \tau_{d_n}>_g.
\end{eqnarray}
The first equation in (\ref{strdileq in 3}) is the string equation, and the second equation in (\ref{strdileq in 3}) is the dilaton equation.
\par The two constraints, $L_{-1} exp(F)=0$ and $L_0 exp(F)=0$, precisely correspond to the string equation and the dilaton equation (\ref{strdileq in 3}) \cite{Mirzakhani2007int}.

\subsection{Large-$g$ asymptotic intersection numbers and Virasoro constraints}
\label{subsec2.2}
We provide evidence that the large-$g$ asymptotic volumes (\ref{asymptoticVgn in 2.1}) yield valid approximate volumes from the perspective of the Witten conjecture. This involves several steps: first, we compute the intersection numbers from the asymptotic volumes (\ref{asymptoticVgn in 2.1}).
\par As discussed in \cite{Dijkgraaf2018}, the following equality holds owing to a result in \cite{Mirzakhani2007int}:  
\begin{eqnarray}
\label{rel volume and int number in 3}
V_{g,n}(b_1, \ldots, b_n) = &  \int_{\overline{\mathcal{M}}_{g,n}} {\rm exp}(2\pi^2\kappa_1+\frac{1}{2}\sum_{i=1}^n b_i^2 \psi_i) \\ \nonumber
= & \sum_{3g-3+n\ge \sum_{i=1}^n m_i\ge 0} \frac{(2\pi^2)^{3g-3+n-\sum_{i=1}^n m_i}}{(3g-3+n-\sum_{i=1}^n m_i)!\prod_{i=1}^n m_i!}\, \prod_{i=1}^n (\frac{b_i^2}{2})^{m_i}\, <\kappa_1^{3g-3+n-\sum_{i=1}^n m_i} \, \prod_{i=1}^n \psi_i^{m_i}>.
\end{eqnarray}
\footnote{$\kappa_1$ denotes the first Miller--Morita--Mumford class, which is cohomologous to the Weil--Petersson symplectic form $\omega$ times $\frac{1}{2\pi^2}$, that is, $\kappa_1=\frac{\omega}{2\pi^2}$ \cite{Wolpert1983, Wolpert1986}.} As mentioned in the introduction, we used $\mathcal{L}_i$ to represent the cotangent spaces to the marked points in a closed Reimann surface of genus $g$, and we used $\psi_i$ to denote the first Chern class of the cotangent space, $\mathcal{L}_i$, namely $\psi_i=c_1(\mathcal{L}_i)$. The $m_i$s are non-negative integers that satisfy condition $3g-3+n\ge \sum_{i=1}^n m_i\ge 0$. $b_1, \ldots, b_n$ denote the lengths of the geodesic boundaries. Utilizing equality (\ref{rel volume and int number in 3}) and based on the large $g$ asymptotics of the Weil--Petersson volumes deduced in \cite{Kimura2008}, we compute the intersection numbers of the cohomology classes associated with the line bundles on the moduli of the Riemann surfaces with boundaries, under conditions $g>>b_i$, $i=1, \ldots, n$, and $g>>1$. This is the capacity to which our argument applies in this section. 
\par Applying expression (\ref{asymptoticVgn in 2.1}) for the large $g$ asymptotic of the Weil--Petersson volume obtained in \cite{Kimura2008} into equality (\ref{rel volume and int number in 3}), we obtain the following relation \footnote{Relation (\ref{equality int num in 3}) holds only under $g>>b_i$, $i=1, \ldots, n$, and $g>>1$. The right-hand side contains finitely many terms in $b_i$, whereas the left-hand side (LHS) contains infinitely many terms in $b_i$; the relation does not hold when $b_i>>g$.}:
\begin{eqnarray}
\label{equality int num in 3}
\sqrt{\frac{2}{\pi}} 2^n \, (4\pi^2)^{2g+n-3} \, \Gamma(2g+n-\frac{5}{2})\, \prod^n_{i=1} \frac{{\rm sinh}(\frac{b_i}{2})}{b_i} \sim & \\ \nonumber
 \sum_{3g-3+n\ge \sum_{i=1}^n m_i\ge 0} \frac{(2\pi^2)^{3g-3+n-\sum_{i=1}^n m_i}}{(3g-3+n-\sum_{i=1}^n m_i)!\prod_{i=1}^n m_i!}\, \prod_{i=1}^n (\frac{b_i^2}{2})^{m_i}\, <\kappa_1^{3g-3+n-\sum_{i=1}^n m_i} \, \prod_{i=1}^n \psi_i^{m_i}>. &
\end{eqnarray}
Because $\frac{{\rm sinh}(\frac{b}{2})}{b}$ has the following Taylor expansion:
\begin{equation}
\label{Taylor exp in 3}
\frac{{\rm sinh}(\frac{b}{2})}{b} = \sum_{n=0}^\infty \frac{b^{2n}}{2^{2n+1}\cdot (2n+1)!},
\end{equation}
the coefficient of $\prod_{i=1}^n b_i^{2m_i}$ on the LHS of (\ref{equality int num in 3}) can be determined from (\ref{Taylor exp in 3}). By comparing the coefficients of $\prod_{i=1}^n b_i^{2m_i}$ on both sides of (\ref{equality int num in 3}), we determine the intersection number as follows under $g>>b_i$, $i=1, \ldots, n$, and $g>>1$:
\begin{equation}
\label{computed int num in 3}
<\kappa_1^{3g-3+n-\sum_{i=1}^n m_i} \, \prod_{i=1}^n \psi_i^{m_i}>^{asymp.} \sim \sqrt{\frac{2}{\pi}} 2^{g+n-3} \, \pi^{2\sum_{i=1}^n m_i-2g} \, \cdot \Gamma(2g+n-\frac{5}{2})\, \frac{(3g-3+n-\sum_{i=1}^n m_i)! \prod_{i=1}^n m_i!}{\prod_{i=1}^n (2m_i+1)!}.
\end{equation}
\par Thus, we obtain the intersection numbers with $n$ number of boundaries where $n\ge 2$ under conditions $g>>b_i$, $i=1, \ldots, n$, and $g>>1$. We placed $asymp.$ in the superscript to emphasize that the intersection numbers are asymptotic approximations obtained using the large-$g$ asymptotic volumes (\ref{asymptoticVgn in 2.1}).

\par The next step is to deduce the asymptotic intersection indices, $<\tau_{d_1}, \ldots, \tau_{d_n}>_g^{asymp.}$, from the asymptotic intersection numbers (\ref{computed int num in 3}).	We placed $asymp.$ in the superscript to emphasize that these values are asymptotic approximations obtained using (\ref{computed int num in 3}). Considering the situation where $\sum_{i=1}^n m_i=3g-3+n$ in the equation (\ref{computed int num in 3}), we obtain the large-$g$ asymptotic of the intersection index, $<\tau_{d_1}, \ldots, \tau_{d_n}>_g^{asymp.}$, as follows:
\begin{equation}
\label{asymp int index in 3}
<\tau_{d_1}, \ldots, \tau_{d_n}>_g^{asymp.}=<\prod_{i=1}^n \psi_i^{d_i}>^{asymp.} \sim \sqrt{\frac{2}{\pi}} 2^{g+n-3} \, \pi^{2(2g-3+n)} \, \cdot \Gamma(2g+n-\frac{5}{2})\, \frac{\prod_{i=1}^n d_i!}{\prod_{i=1}^n (2d_i+1)!}.
\end{equation} 
\par Following the mathematical formulation of the Witten conjecture, we define the asymptotic generating function of the intersection indices (\ref{asymp int index in 3}) obtained from the asymptotic intersection numbers (\ref{computed int num in 3}). We denote the generating function of the large-$g$ asymptotic intersection indices $<\tau_{d_1}, \ldots, \tau_{d_n}>_g^{asymp.}$ as $\til{F}$. Namely, $\til{F}$ is defined as:
\begin{equation}
\label{asymp generating fn}
\til{F}=\sum_{g=0}^{\infty} \til{F}_g \zeta^{2g-2},
\end{equation}
where
\begin{equation}
\label{asymptotic sum of Fg in 3}
\til{F}_g(t_0, t_1, \ldots) = \sum_{\{d_i\}} \, \prod_{l>0} \frac{t_l^{n_l}}{n_l!} <\prod \tau_{d_i}>_g^{asymp.}.
\end{equation}
(The values of the large-$g$ asymptotic intersection indices $<\prod \tau_{d_i} >_g^{asymp.}$ in the sum (\ref{asymptotic sum of Fg in 3}) are given as (\ref{asymp int index in 3}).)
As we stated in the introduction, we refer to the resulting generating function $\til{F}$ (\ref{asymp generating fn}) as the asymptotic generating function. 
\par As we stated earlier in the introduction, if the asymptotic volumes (\ref{asymptoticVgn in 2.1}) yield correct approximate formulas, then the asymptotic generating should possess a symmetry analogous to that of the generating function $F$ (\ref{equation generating function}). To be concrete, we expect that the partition function of the asymptotic generating function $\til{F}$, $exp(\til{F})$, is annihilated by a series of partial differential operators analogous to the operators $L_n$ in the Virasoro constraints (\ref{eq Virasoro constraints}). Because the generating function $\til{F}$ is defined from the asymptotic volumes (\ref{asymptoticVgn in 2.1}), it is unlikely that the asymptotic generating function $\til{F}$ satisfies the exact Virasoro constraints. 
\par It seems a considerably difficult problem to determine all partial differential operators that annihilate the asymptotic generating function $\til{F}$. However, we provide supporting evidence that such partial differential operators analogous to the operators $L_n$ in the Virasoro constraints indeed exist. Presenting this evidence is the last step. 
\par As supporting evidence for the asymptotic generating function $\til{F}$ satisfying constraints analogous to the Virasoro constraints, we demonstrate explicitly that the asymptotic intersection indices (\ref{asymp int index in 3}) deduced from the intersection numbers (\ref{computed int num in 3}) satisfy variants of the string and dilaton equations. 
\par It is known that the constraints where the partition function $exp(F)$ of the generating function $F$ is annihilated by the operators $L_{-1}$ and $L_0$, $L_{-1}\, exp(F)=0$ and $L_0\, exp(F)=0$, are equivalent to the string and dilaton equations (\ref{strdileq in 3}) as we explained at the end of section \ref{subsec2.1}. Therefore, when the asymptotic intersection indices (\ref{asymp int index in 3}) satisfy variants of the string and dilaton equations, that implies that there exist variants of the differential operators $L_{-1}$ and $L_0$, which we denote as $\til{L}_{-1}$ and $\til{L}_0$, and the partition function of the asymptotic generating function $\til{F}$ (\ref{asymp generating fn}), $exp(\til{F})$, is annihilated by the operators $\til{L}_{-1}$ and $\til{L}_0$. 

\par Now, as the last step, we demonstrate that the asymptotic intersection indices (\ref{asymp int index in 3}) satisfy the variants of of the string and dilaton equations. Because the large-$g$ asymptotic intersection numbers (\ref{computed int num in 3}) are obtained as approximate values that are valid in the region with the large genus $g$, we do not expect that the intersection indices (\ref{asymp int index in 3}) obtained from the asymptotic expressions (\ref{computed int num in 3}) exactly satisfy the two equations (\ref{strdileq in 3}). Still, after some computations, we confirm that they satisfy the string and dilaton equations up to finite factors.
\par  We compare the ratio of the intersection indices $<\tau_1, \tau_{d_1}, \ldots, \tau_{d_n}>_g^{asymp.} $ to $(2g-2+n) <\tau_{d_1}, \ldots, \tau_{d_n}>_g^{asymp.}$, and the ratio $<\tau_0, \tau_{d_1}, \ldots, \tau_{d_n}>_g^{asymp.} $ to $\sum_{d_i\ne 0} <\tau_{d_1}, \ldots, \tau_{d_i-1}, \ldots>_g^{asymp.} $. The special intersection index $<\tau_{d_1}, \ldots, \tau_{d_n}, \tau_{d_{n+1}}>_g$, where $d_{n+1}=1$, is used to obtain the intersection index $<\tau_1, \tau_{d_1}, \ldots, \tau_{d_n}>_g$. Therefore, we obtain the intersection index $<\tau_1, \tau_{d_1}, \ldots, \tau_{d_n}>_g^{asymp.}$ as follows:
\begin{equation}
<\tau_1, \tau_{d_1}, \ldots, \tau_{d_n}>_g^{asymp.} \sim \sqrt{\frac{2}{\pi}} 2^{g+n-2} \, \pi^{2(2g-2+n)} \, \cdot \Gamma(2g+n-\frac{3}{2})\, \frac{\prod_{i=1}^n d_i!}{3!\, \prod_{i=1}^n (2d_i+1)!}.
\end{equation}
Therefore, we deduce that the ratio $<\tau_1, \tau_{d_1}, \ldots, \tau_{d_n}>_g^{asymp.}$ to $(2g-2+n) <\tau_{d_1}, \ldots, \tau_{d_n}>_g^{asymp.}$ can be expressed as follows:
\begin{equation}
\label{ratio string in 3}
\frac{<\tau_1, \tau_{d_1}, \ldots, \tau_{d_n}>_g^{asymp.}}{(2g-2+n) <\tau_{d_1}, \ldots, \tau_{d_n}>_g^{asymp.}}=\frac{2\pi^2}{3!}\cdot \frac{2g+n-\frac{5}{2}}{2g+n-2}.
\end{equation}
The computational result (\ref{ratio string in 3}) revealed that the ratio is not 1; however, the ratio (\ref{ratio string in 3}) tends toward $\frac{\pi^2}{3}$ as the genus $g$ tends toward infinity. Therefore, we deduce that the large-$g$ asymptotic intersection indices computed from the asymptotic expression (\ref{computed int num in 3}) satisfy a variant of the dilaton equation when the genus $g$ is large:
\begin{equation}
\label{variant str eq in 3}
<\tau_1, \tau_{d_1}, \ldots, \tau_{d_n}>_g^{asymp.} = \frac{\pi^2}{3} (2 g-2 +n) <\tau_{d_1}, \ldots, \tau_{d_n}>_g^{asymp.} 
\end{equation}
\par Next, we demonstrate that the large-$g$ asymptotic intersection indices satisfy a variant of the string equation. The LHS term of the string equation, $<\tau_0, \tau_{d_1}, \ldots, \tau_{d_n}>_g $, is a special case of the intersection index $<\tau_{d_1}, \ldots, \tau_{d_n}, \tau_{d_n+1}>_g $, where $d_{n+1}=0$. Because there are $n+1$ marked points for this equation, we have $\sum_{i=1}^{n+1} d_i=3g-3+n+1=3g-2+n$, and for the special situation $d_{n+1}=0$, we have $\sum_{i=1}^n d_i=3g-2+n$. Large-$g$ asymptotic intersection index $<\tau_0, \tau_{d_1}, \ldots, \tau_{d_n}>_g^{asymp.} $ is then expressed as follows:
\begin{equation}
<\tau_0, \tau_{d_1}, \ldots, \tau_{d_n}>_g^{asymp.} \sim \sqrt{\frac{2}{\pi}} 2^{g+n-2} \, \pi^{2(2g-2+n)} \, \cdot \Gamma(2g+n-\frac{3}{2})\, \frac{\prod_{i=1}^n d_i!}{\prod_{i=1}^n (2d_i+1)!}.
\end{equation}
For simplicity, we focus on the cases where $d_i$, $i=1, \ldots, n$, are nonzero. The large-$g$ asymptotic intersection index $<\tau_{d_1}, \ldots, \tau_{d_i-1}, \ldots>_g^{asymp.}$ is expressed as follows:
\begin{equation}
<\tau_{d_1}, \ldots, \tau_{d_i-1}, \ldots>_g^{asymp.} \sim \sqrt{\frac{2}{\pi}} 2^{g+n-3} \, \pi^{2(2g-3+n)} \, \cdot \Gamma(2g+n-\frac{5}{2})\, \frac{\prod_{j\ne i} d_j!\cdot (d_i-1)!}{\prod_{j\ne i} (2d_j+1)!\cdot (2d_i-1)!}.
\end{equation}
We then find that the sum $\sum_{d_i\ne 0} <\tau_{d_1}, \ldots, \tau_{d_i-1}, \ldots>_g^{asymp.} $ is expressed as follows:
\begin{eqnarray}
\sum_{d_i\ne 0} <\tau_{d_1}, \ldots, \tau_{d_i-1}, \ldots>_g^{asymp.} & \sim  \sqrt{\frac{2}{\pi}} 2^{g+n-3}  \pi^{2(2g-3+n)} \,  \Gamma(2g+n-\frac{5}{2})\,\sum_{i=1}^n \frac{\prod_{j\ne i} d_j!\cdot (d_i-1)!}{\prod_{j\ne i} (2d_j+1)!\cdot (2d_i-1)!} \\ \nonumber
= & \sqrt{\frac{2}{\pi}} 2^{g+n-3} \, \pi^{2(2g-3+n)} \, \cdot \Gamma(2g+n-\frac{5}{2})\, \frac{\prod_{j=1}^n d_j!}{\prod_{j=1}^n (2d_j+1)!}\cdot  \sum_{i=1}^n 2(2d_i+1) \\ \nonumber
= & \sqrt{\frac{2}{\pi}} 2^{g+n-3} \, \pi^{2(2g-3+n)} \, \cdot \Gamma(2g+n-\frac{5}{2})\, \frac{\prod_{j=1}^n d_j!}{\prod_{j=1}^n (2d_j+1)!}\cdot  2(6g-4+3n).
\end{eqnarray}
Therefore, the ratios $<\tau_0, \tau_{d_1}, \ldots, \tau_{d_n}>_g^{asymp.} $ to $\sum_{d_i\ne 0} <\tau_{d_1}, \ldots, \tau_{d_i-1}, \ldots>_g^{asymp.} $ can be expressed as follows:
\begin{equation}
\label{ratiodil in 3}
\frac{<\tau_0, \tau_{d_1}, \ldots, \tau_{d_n}>_g^{asymp.}}{\sum_{d_i\ne 0} <\tau_{d_1}, \ldots, \tau_{d_i-1}, \ldots>_g^{asymp.}} = 2\pi^2 \cdot \frac{2g+n-\frac{5}{2}}{2(6g-4+3n)}.
\end{equation}
In the limit at which the genus $g$ goes to infinity, the ratio (\ref{ratiodil in 3}) tends toward $\frac{\pi^2}{3}$. Thus, we find that the large-$g$ asymptotic intersection indices deduced from (\ref{computed int num in 3}) satisfy a variant of the string equation in the large-$g$ limit as follows:
\begin{equation}
\label{variant dil eq in 3}
<\tau_0, \tau_{d_1}, \ldots, \tau_{d_n}>_g^{asymp.} = \frac{\pi^2}{3} \sum_{d_i\ne 0} <\tau_{d_1}, \ldots, \tau_{d_i-1}, \ldots>_g^{asymp.} 
\end{equation}

\par In brief, we have shown that the asymptotic intersection indices (\ref{asymp int index in 3}) satisfy variants of the string and dilaton equations as follows:
\begin{eqnarray}
\label{var strdileq}
<\tau_0, \tau_{d_1}, \ldots, \tau_{d_n}>_g^{asymp.} = & \frac{\pi^2}{3} \sum_{d_i\ne 0} <\tau_{d_1}, \ldots, \tau_{d_i-1}, \ldots>_g^{asymp.}, \\ \nonumber
<\tau_1, \tau_{d_1}, \ldots, \tau_{d_n}>_g^{asymp.} = & \frac{\pi^2}{3} (2 g-2 +n) <\tau_{d_1}, \ldots, \tau_{d_n}>_g^{asymp.}.
\end{eqnarray}

\par We observed that the large-$g$ asymptotic intersection indices (\ref{asymp int index in 3}) obtained from the asymptotic intersection numbers (\ref{computed int num in 3}) satisfy variants of dilaton and string equations in the large-$g$ limit, as in (\ref{var strdileq}). These results imply that the partition function of the generating function of the large-$g$ asymptotic intersection indices satisfies the variants of the two constraints $L_{-1} exp(F)=0$ and $L_0 exp(F)=0$. Equations (\ref{var strdileq}) imply that there exist variants of operators $L_0$ and $L_{-1}$, which we denote as $\til{L}_0$ and $\til{L}_{-1}$, and the partition function $exp(\til{F})$ satisfies the constraints $\til{L}_0 exp(\til{F})=0$ and $\til{L}_{-1} exp(\til{F})=0$.
\par These observations lead us to postulate that the large-$g$ asymptotic partition function $exp(\til{F})$ satisfies a variant of Virasoro constraints in the large-$g$ limit. We expect that the variants of Virasoro operators, which we denote as $\til{L}_n$, $n\ge 1$, also exist such that the partition function $exp(\til{F})$ satisfies the constraints $\til{L}_n exp(\til{F})=0$. The construction of all such operators can be undertaken in a future study.

\section{Large-$g$ asymptotic Weil--Petersson volume and higher-order spectral form factor}
\label{sec3}
As another theme, we discuss the higher-order spectral form factor (SFF) in this section. The higher-order SFFs were analyzed in \cite{Mertens2020}. We take an approach that is different from that used in \cite{Mertens2020}, and we study the behaviors of the higher-order SFFs utilizing the large-$g$ asymptotics (\ref{asymptoticVgn in 2.1}). The agreement of our result with the result given in \cite{Mertens2020} provides another evidence supporting the validity of the asymptotic formulas (\ref{asymptoticVgn in 2.1}). 
\par As a first step, we compute the genus-$g$ partition function via the Laplace transform of (\ref{asymptoticVgn in 2.1}). The connected correlation function of JT gravity on a Riemann surface of genus $g$ with $n$ asymptotic boundaries has genus expansion is as follows \cite{SSS2019}: 
\begin{equation}
\label{genus expansion corr in intro}
<Z(\beta_1) \ldots Z(\beta_n)>_c \simeq \sum^\infty _{g=0} Z_{g,n}(\beta_1, \ldots, \beta_n) \cdot e^{(-2g+2-n)S_0}.
\end{equation}
Here, $Z_{g,n}(\beta_1, \ldots, \beta_n)$ denotes the JT path integral for the Riemann surfaces with a topology of genus $g$ with $n$ asymptotic boundaries. The genus-$g$ partition function with $n$ boundaries, $Z_{g,n}(\beta_1, \ldots, \beta_n)$, can be expressed as follows \cite{SSS2019} \footnote{There is a normalization constant, $\alpha$, in the genus-$g$ partition functions \cite{SSS2019}. We adopt the convention to set $\alpha=1$ in this study.} :
\begin{equation}
Z_{g,n}(\beta_1, \ldots, \beta_n) = \prod^n_{i=1} \int^\infty_0 b_idb_i \, V_{g,n}(b_1, \ldots, b_n) \prod^n_{j=1} Z_{\rm Sch}^{\rm trumpet}(\beta_j, b_j),
\end{equation}
where $Z_{\rm Sch}^{\rm trumpet}(\beta, b)$ denotes the contribution from the path integral over the trumpet connecting an asymptotic wiggly boundary and a geodesic boundary of length $b$. $V_{g,n}(b_1, \ldots, b_n)$ denotes the Weil--Petersson volume of the moduli of the Riemann surfaces of genus $g$ with $n$ geodesic boundaries. $b_1, \ldots, b_n$ denote the lengths of the geodesic boundaries. The contribution from the trumpet, $Z_{\rm Sch}^{\rm trumpet}$, was computed in \cite{SSS2019}. As stated previously, one can compute the correlation function with $n$ boundaries by evaluating the values of the Weil--Petersson volumes, $V_{g,n}(b_1, \ldots, b_n)$. The contribution, $Z_{\rm Sch}^{\rm trumpet}(\beta, b)$, was computed in \cite{SSS2019} as $\sqrt{\frac{\gamma}{2\pi \beta}}e^{-\frac{\gamma b^2}{2\beta}}$. 
\par Utilizing expression (\ref{asymptoticVgn in 2.1}), after a small computation we find that the genus-$g$ partition function with $n$ boundaries can be derived as follows when $g>>1$: 
\begin{equation}
\begin{aligned}
Z_{g,n}(\beta_1, \ldots, \beta_n) = & \sqrt{\frac{2}{\pi}} 2^n \, (4\pi^2)^{2g+n-3} \, \Gamma(2g+n-\frac{5}{2})\, \prod_{i=1}^n\sqrt{\frac{\gamma}{2\pi \beta_i}} \, \prod_{j=1}^n \int^\infty_0 db_j {\rm sinh}(\frac{b_j}{2})e^{-\frac{\gamma b_j^2}{2\beta_j}} \\
 = & \sqrt{\frac{2}{\pi}} \, (4\pi^2)^{2g+n-3} \, \Gamma(2g+n-\frac{5}{2})\, e^{\frac{\sum_{i=1}^n \beta_i}{8\gamma}} \prod_{j=1}^n {\rm erf} (\sqrt{\frac{\beta_j}{8\gamma}}).
\end{aligned}
\end{equation}
The resulting function is the product of a coefficient that depends on genus $g$, an exponential function, and the error functions.
\par We deduce that the higher-genus contributions to the correlation function with $n$ boundaries are given as follows:
\begin{equation}
\label{sum higher genus with n boundaries in 2.1}
 \sqrt{\frac{2}{\pi}}  \, e^{\frac{\sum_{i=1}^n \beta_i}{8\gamma}} \prod_{j=1}^n {\rm erf} (\sqrt{\frac{\beta_j}{8\gamma}}) \cdot \sum_{g>>1} (4\pi^2)^{2g+n-3} \, \Gamma(2g+n-\frac{5}{2})\cdot e^{(-2g+2-n)S_0}.
\end{equation}
\par From the computed correlation functions, we learn that the nonperturbative correction takes the form of $e^{-\frac{e^{S_0}}{4\pi^2}}$. This agrees with the results of the correction and the effect of the ZZ brane \cite{Zamolodchikov2001}, as discussed in \cite{Okuyama2019}.
\par The method applies to cases with two or more boundaries, i.e. $n\ge 2$. The higher-genus contributions to the correlation function with one boundary can also be deduced in a similar manner using the large-$g$ asymptotic of the Weil--Petersson volume with one boundary, $V_{g,1}(b)$, as deduced in \cite{SSS2019}.

\vspace{5mm}

\par Now, we compute the higher-order SFFs. Substituting $\beta\pm iT$ for $\beta$ in (\ref{sum higher genus with n boundaries in 2.1}), we obtained the higher-genus contributions to the higher-order spectral form factor $<\prod_n Z(\beta+i T) Z(\beta-i T)>_c$ as
\begin{equation}
\label{higher SFF in 2.1}
\sqrt{\frac{2}{\pi}}  \, e^{\frac{2n\beta}{8\gamma}} ({\rm erf} (\sqrt{\frac{\beta+i T}{8\gamma}}))^n ({\rm erf} (\sqrt{\frac{\beta-i T}{8\gamma}}))^n \cdot \sum_{g>>1} (4\pi^2)^{2g+2n-3} \, \Gamma(2g+2n-\frac{5}{2})\cdot e^{(-2g+2-2n)S_0}.
\end{equation}
\par The leading-order term in the genus expansion of the higher-order spectral form factor $<\prod_n Z(\beta+i T) Z(\beta-i T)>_c$ was computed in \cite{Mertens2020}, and the result in \cite{Mertens2020} revealed that the leading-order term is proportional to $(\beta^2+T^2)^{\frac{n}{2}}$. From the obtained higher-genus contribution (\ref{higher SFF in 2.1}), we determine that this property is persistent in the higher-genus terms when $\beta$ is small because ${\rm erf}(z) \sim \frac{2}{\sqrt{\pi}}\,z$ when $z$ is small, and $e^{\frac{2n\beta}{8\gamma}} ({\rm erf} (\sqrt{\frac{\beta+i T}{8\gamma}}))^n ({\rm erf} (\sqrt{\frac{\beta-i T}{8\gamma}}))^n$ in (\ref{higher SFF in 2.1}) is approximated by $\frac{1}{(8\gamma)^n}\frac{2^{2n}}{\pi^n}\, (\beta^2+T^2)^{\frac{n}{2}}$ when $\beta$ is small. Thus, the higher-genus terms are still proportional to $(\beta^2+T^2)^{\frac{n}{2}}$ when $\beta$ is small.
\par The behavior of the higher-order SFF that we deduced using (\ref{asymptoticVgn in 2.1}) agrees with the result in \cite{Mertens2020}. The higher-order SFF being proportional to $(\beta^2+T^2)^{\frac{n}{2}}$, from the perspective of our approach, originates from a property of the formula (\ref{asymptoticVgn in 2.1}) given as product of hyperbolic functions, $\frac{{\rm sinh}(\frac{b_i}{2})}{b_i}$. The agreement provides another supporting evidence for the asymptotic formula (\ref{asymptoticVgn in 2.1}).

\section{Concluding remarks}
\label{sec4}	
We examined whether the large-$g$ asymptotic Weil--Petersson volumes (\ref{asymptoticVgn in 2.1}) are valid. We used two approaches to examine the validity of the formula. The volume formulas are relevant to computations of the path integrals in JT gravity. 
\par In the first approach we used techniques in the Witten conjecture, and we observed that the asymptotic intersection indices (\ref{asymp int index in 3}) computed in section \ref{subsec2.2} satisfy variants of the string and dilaton equations (\ref{var strdileq}) in the large-$g$ limit. The string and dilaton equations precisely correspond to the two constraints, $L_0 \, exp(F)=0$ and $L_{-1} \, exp(F)=0$ in the Virasoro constraints (\ref{eq Virasoro constraints}). These correspondences imply that because the asymptotic intersection indices deduced in section \ref{sec3} satisfy variants of string and dilaton equations, the partition function $exp(\til{F})$ of the generating function $\til{F}$ of the asymptotic intersection indices deduced in this study satisfies a variant of the two constraints. 
\par This seems to suggest that the partition function $exp(\til{F})$ of the generating function of the asymptotic intersection indices satisfies a variant of the Virasoro constraints. 
\par As the other approach, we investigated the behavior of the higher-order SFFs to examine the asymptotic volume formulas (\ref{asymptoticVgn in 2.1}). We deduced the behaviors of the higher-order SFFs from the formulas (\ref{asymptoticVgn in 2.1}), and we confirmed that the resulting behaviors agreed with those in \cite{Mertens2020}. This result provided additional supporting evidence for the formulas (\ref{asymptoticVgn in 2.1}). 
\par The results deduced in this study suggest, to some degree, that the large-$g$ asymptotic Weil--Petersson volume formulas (\ref{asymptoticVgn in 2.1}) yield plausible estimations.

\section*{Acknowledgments}

We would like to thank Shun'ya Mizoguchi and Kazuhiro Sakai for discussions.	

\appendix
\section{Resolvents on the matrix-integral side}
\label{appendixa}
The correlation functions of the resolvents in the double-scaled matrix integral as dual of JT gravity have the following genus expansion \cite{SSS2019}:
\begin{equation}
\label{resolv genus exp in 2.2}
<R(E_1)\ldots R(E_n)>_c = \sum^{\infty}_{g=0} R_{g,n}(E_1, \ldots, E_n)\cdot e^{(-2g-n+2)S_0}.
\end{equation}
Genus expansion (\ref{resolv genus exp in 2.2}) defines the multi-resolvent correlators, $R_{g,n}$, as discussed in \cite{SSS2019}. We compute the multi-resolvent correlators, $R_{g,n}$, when genus $g$ is large. This yields the large-genus contributions to the correlation functions of the resolvents, $<R(E_1)\ldots R(E_n)>_c$. 
\par Functions $W_{g,n}$ are defined as follows, as discussed in \cite{Eynard2014, SSS2019} :
\begin{equation}
\label{def W in 2.2}
W_{g,n}(z_1, \ldots, z_n) = (-2)^n\cdot \prod^n_{i=1} z_i \cdot R_{g,n}(-z_1^2, \ldots, -z_n^2).
\end{equation}
As proved in \cite{Eynard2007}, function $W_{g,n}$ is given in terms of the Weil--Petersson volume, $V_{g,n}$, as follows:
\begin{equation}
\label{rel W and WP vol in 2.2}
W_{g,n}(z_1, \ldots, z_n) = \prod^n_{i=1} \int_0^{\infty} b_i\, db_i \, e^{-b_i z_i} \, \cdot V_{g,n}(b_1, \ldots, b_n).
\end{equation}
For the region where $g>>1$, using the equation (\ref{rel W and WP vol in 2.2}), we can compute $W_{g,n}$ by utilizing the large-$g$ asymptotic of the Weil--Petersson volume, $V_{g,n}$, deduced in \cite{Kimura2008}. Based on the definition of $W_{g,n}$ (\ref{def W in 2.2}), the multi-resolvent correlator, $R_{g,n}$, is deduced from the computed $W_{g,n}$. 
\par Applying the large-$g$ asymptotic Weil--Petersson volume (\ref{asymptoticVgn in 2.1}) into (\ref{rel W and WP vol in 2.2}), we obtain $W_{g,n}$ as follows:
\begin{eqnarray}
\label{int W in 2.2}
W_{g,n}(z_1, \ldots, z_n) & \sim \sqrt{\frac{2}{\pi}} 2^n \, (4\pi^2)^{2g+n-3} \, \Gamma(2g+n-\frac{5}{2})\, \prod^n_{i=1} \int^{\infty}_0 db_i {\rm sinh}(\frac{b_i}{2})\, e^{-b_i z_i} \\ \nonumber
& = 2^n \sqrt{\frac{2}{\pi}}  \, (4\pi^2)^{2g+n-3} \, \Gamma(2g+n-\frac{5}{2})\, \frac{(-2)^n}{\prod^n_{i=1}\big( 4(-z_i^2)+1 \big)}.
\end{eqnarray}
\par According to this result, in region $g>>1$, the multi-resolvent correlator, $R_{g,n}$, is determined as follows:
\begin{equation}
R_{g,n}(E_1, \ldots, E_n) \sim 2^n \sqrt{\frac{2}{\pi}}  \, (4\pi^2)^{2g+n-3} \, \Gamma(2g+n-\frac{5}{2})\, \frac{1}{\prod^n_{i=1}(4E_i+1)} \frac{1}{\sqrt{(-1)^n \prod^n_{i=1} E_i}}.
\end{equation}
Therefore, we deduce that the higher-genus contributions to the correlation functions of the resolvents of the double-scaled matrix integral with $n$ boundaries can be given as follows:
\begin{equation}
\sum_{g>>1} 2^n \sqrt{\frac{2}{\pi}}  \, (4\pi^2)^{2g+n-3} \, \Gamma(2g+n-\frac{5}{2})\, \frac{1}{\prod^n_{i=1}(4E_i+1)} \frac{1}{\sqrt{(-1)^n \prod^n_{i=1} E_i}} \cdot e^{(-2g-n+2)S_0}.
\end{equation}
\par The described method applies to cases with two or more boundaries. Using the large-$g$ asymptotic of the Weil--Petersson volume with one boundary, $V_{g,1}(b)$, as deduced in \cite{SSS2019}, we can obtain the higher-genus contributions to the resolvent with one boundary.

\end{document}